\documentstyle[11pt]{article}

\begin{document}

\title{{\bf Nonadiabatic approach to dimerization gap and optical absorption
coefficient of the Su-Schrieffer-Heeger model}}
\author{Qin Wang$^{{\rm a,b}}$, \ Hong Chen$^{{\rm a}}$ \ and \
Hang Zheng$^{{\rm b}}$ \\
{\normalsize $^{{\rm a}}${\em Pohl Institute of Solid State Physics, Tongji
University, }}\\
{\normalsize {\em Shanghai 200092, People's Republic of China}}\\
{\normalsize {\em $^{{\rm b}}$Department of Applied Physics, Shanghai Jiao
Tong University, }}\\
{\normalsize {\em Shanghai 200030, People's Republic of China}}}
\date{}
\maketitle

\begin{abstract}
An analytical nonadiabatic approach has been developed to study the
dimerization gap and the optical absorption coefficient of the
Su-Schrieffer-Heeger model where the electrons interact with dispersive
quantum phonons. By investigating quantitatively the effects of quantum
phonon fluctuations on the gap order and the optical responses in this
system, we show that the dimerization gap is much more reduced by the
quantum lattice fluctuations than the optical absorption coefficient is. The
calculated optical absorption coefficient and the density of states do not
have the inverse-square-root singularity, but have a peak above the gap edge
and there exist a significant tail below the peak. The peak of optical
absorption spectrum is not directly corresponding to the dimerized gap. Our
results of the optical absorption coefficient agree well with those of the
experiments in both the shape and the peak position of the optical
absorption spectrum. \newline
PACS number(s): 71.38.-k; 71.20.Rv; 78.20.Ci; 71.45.Lr\newline
\end{abstract}

\newpage

\section{Introduction}

Since the pioneering work of Su, Schrieffer and Heeger (SSH) \cite{ra1}, the
SSH model has been widely used to investigate the properties of conducting
polymers, especially polyacetylene [(CH)$_{x}$] \cite{ra2}. At half-filling,
this system, as many quasi-one-dimensional materials, exhibits an
instability against a periodic lattice distortion due to the Peierls
dynamics. Because of a twofold degenerate ground state, this system allows
for the soliton excitation with an associated electronic state at the center
of the energy gap \cite{ra3,ra4}. While considerable effort has been made to
understand excitation energies of solitons and polarons in these compounds,
it is theoretically and experimentally significant to develop an accepted
theory for their optical absorption spectrum. The recent exploitations of
this model in studying the properties of nanotubes \cite{rb1,rb2,rb3} renew
further interesting of theoretical studies on the intrinsic physics of this
model. In the previous works on conducting polymers, along with the
expanding of measurements, theoretical attempts have been made to study the
energy gap \cite{ra5}, low-lying excitation \cite{ra3,ra4,ra6}, and optical
response \cite{ra3,ra7} of this system. However, there exist discrepancies
between theories and experiments. (i) Within the adiabatic approximation,
the calculated optical absorption coefficient of the perfectly dimerized
lattice has an inverse-square-root edge singularity at $\omega =2\Delta $
and there is no absorption inside the gap. But the observed optical
absorption spectrum of polyacetylene is quite different \cite{ra8,ra9,ra10}.
The singularity is absent, and there is a significant tail below the
maximum. (ii) The energy gap deduced from the absorption edge is smaller
than the activation energy of the dc conductivity \cite{ra11,ra12}.

In (CH)$_{x}$ the zero-point motion of the lattice $\delta u$ (=0.03 \cite%
{ra12}) is comparable to the amplitude of the lattice Peierls distortion $%
u_{0}$ (=0.03 \cite{ra13} or 0.035 \cite{ra14}). Although it has been stated
that the quantum lattice fluctuations cause a slightly small reduction of
the dimerization order parameter \cite{ra15,ra16,ra17}, if $\delta u\sim
u_{0}$, as was pointed out in Ref. \cite{ra2,ra12}, one might wonder why
there should be a clearly defined gap in the spectrum at all and expect the
optical absorption spectrum to peak at the correspondingly renormalized gap
edge. In nonadiabatic case, the phonon frequency $\omega _{\pi }$ is finite.
Generally speaking, the nonadiabatic effect will suppress the state order
parameters of the system \cite{ra16}. As the density-of-states (DOS) of
electrons is concerned, the results of adiabatic approximations also have
inverse-square-root singularity at the gap edge. By considering the
nonadiabatic effect, the singularity may disappear \cite{ra18}. The
influence of the phonon frequency on DOS, the optical-absorption spectrum,
and other order parameters in the range from $\omega _{\pi }=0$ to $\omega
_{\pi }\rightarrow \infty $ should be studied for understanding the physics
of electron-phonon interactions in nonadiabatic case.

The theoretical analysis of the coupled electron-phonon system for all
values of the ionic mass and the electron-phonon coupling constant is a very
difficult problem, especially when the quantum lattice fluctuation and the
lattice dynamic (quantum phonon) effects are taken into account. This
problem has been investigated by using the Monte Carlo simulation \cite%
{ra15,ra16}, perturbation calculation \cite{ra19}, Green's-function
technique \cite{ra7}, renormalization group analysis \cite{ra20,ra21},
variational method of the squeezed-polaron wave-function \cite{ra22}, etc.
In a resent work \cite{ra23}, this problem was studied by considering the
spinless Holstein model and obtained some interesting results. In this
paper, by considering SSH model, the more realistic case with spin-$\frac{1}{%
2}$ electrons, we focus on the properties of the effects of quantum lattice
fluctuations on the dimerization energy gap and the optical-responses of the
system with the view of understanding the measurement evidences and of
interpreting why and how the dimerization energy gap is smaller than the
optical absorption gap. Our analyses show that at finite phonon frequency,
the peak position of optical absorption spectrum is not directly
corresponding to the dimerized gap. The dimerization gap is much more
reduced by the quantum lattice fluctuations than the optical absorption
coefficient is. The effects of quantum lattice fluctuations on the
dimerization gap and on the optical gap are essentially different,
especially when phonon frequency $\omega _{\pi }$ is small. The reason for
this is discussed and the relationship between the peak position of optical
absorption spectrum and the true dimerization gap is obtained quantitatively.

\section{Model}

We start from the SSH model \cite{ra1} 
\begin{equation}
H=-\sum_{l,s}[t+\alpha (u_{l}-u_{l+1})](c_{l,s}^{\dag
}c_{l+1,s}+c_{l+1,s}^{\dag }c_{l,s})+\sum_{l}\left[ \frac{1}{2M}P_{l}^{2}+%
\frac{1}{2}K(u_{l}-u_{l+1})^{2}\right] .  \label{eq:hamilt}
\end{equation}
The notations in it are as usual \cite{ra2} (throughout this paper, we set $%
\hbar =k_{B}=1$).

In Hamiltonian (\ref{eq:hamilt}) the operators of lattice modes, $u_{l}$ and 
$p_{l}$, can be expanded by the phonon creation and annihilation operators $%
b_{q}^{\dag }$ and $b_{q}$, and after Fourier transformation to momentum
space, the $H$ becomes 
\begin{equation}
H=\sum_{k,s}\epsilon _{k}c_{k,s}^{\dag }c_{k,s}+\sum_{q}\omega _{q}\left(
b_{q}^{\dag }b_{q}+{\frac{{1}}{{2}}}\right) +\frac{1}{\sqrt{N}}%
\sum_{q,k,s}g(k+q,k)(b_{-q}^{\dag }+b_{q})c_{k+q,s}^{\dag }c_{k,s},
\label{eq:hamilt2}
\end{equation}
where $\epsilon _{k}=-2t\cos k$ is the bare band structure, $N$ is the total
number of sites. The phonon frequency 
\begin{equation}
\omega _{q}=\sqrt{(4K/M)\sin ^{2}\frac{q}{2}},
\end{equation}
and the coupling function 
\begin{equation}
g(k+q,k)=i2\alpha \sqrt{1/(2M\omega _{q})}[\sin (k+q)-\sin k].
\end{equation}
In the SSH model, the electron couples to the difference between the phonon
amplitudes on the two neighboring sites and the form of the lattice
vibration energy leads the phonons to be dispersive.

Among the models for one-dimensional systems the Holstein model \cite{refb2}
and SSH model are the two typical electron-phonon coupling Hamiltonian
studied by many previous authors. The Holstein model is for the on-site
electron-phonon interaction with dispersionless phonons coupled with
electron density operator and has played a central role in the polaron
problem, while the SSH model is for the on-bond electron-phonon interaction
and has been successful in describing the essential physics of conducting
polymers. Phonon dispersion and different structures of the coupling terms
lead to important differences in the physics of the two models. For the
Holstein model, although numerical simulation \cite{refb3} has been
performed in confirmation of the supposition that it is likely the Peierls
gap is more reduced by the quantum lattice fluctuations than the
dimerization or the charge-density-wave amplitude, in other words, the
quantum lattice fluctuations have a much strong effect on the Peierls gap
than on the amplitude of the Peierls distortion, due to the difficulty in
preparing practicably pure sample of molecular crystal materials, few of
observation data are available for the comparison with theoretical results.
Other models somewhat intermediate between the SSH and Holstein models, such
as the model for $MX$ chain materials \cite{refb4} and the model for
electron-libron coupling in Polyaniline \cite{refb5}, because of their
complex forms with Coulomb interactions and steric potential accounting for
the ring torsion angle, will make the presentation of our analysis to be
more prolix and inexplicit. The enormous measurement data of optical
absorption spectrum in conducting polymers and the relatively simple form of
SSH model render it the favorable system for our investigation into the
nonadiabatic effect on the dimerization gap and the optical absorption
spectrum of the electron-phonon coupling systems. The previous researches on
this problem in SSH model by using a static Gaussian-random potential method
to model the lattice disorder field \cite{ra12,refb6} and a
configurational-coordinate method to treat the anharmonic lattice
fluctuations \cite{ra27} were phenomenological theories and have not
obtained the true Peierls dimerization gap. It is the purpose of this paper
to develop an nonadiabatic approach to study analytically this problem
simply from the model Hamiltonian.

\section{Effective Hamiltonian}

The electron-phonon coupling and the Peierls dimerization are two main
respects in our studying on the SSH model. First, in order to take into
account the electron-phonon coupling, an unitary transformation is applied
to $H$, 
\begin{equation}
H^{\prime}=\exp (S)H\exp (-S),
\end{equation}
where the generator $S$ is 
\begin{equation}
S=\frac{1}{\sqrt{N}}\sum_{q,k,s}\frac{g(k+q,k)}{\omega _{q}}\delta
(k+q,k)(b_{-q}^{\dag }-b_{q})c_{k+q,s}^{\dag }c_{k,s}.
\end{equation}
Here, we introduce a function $\delta (k^{\prime},k)$ which is a function of
the energies of incoming and outgoing electrons in the electron-phonon
scattering process and it's form will be determined later. We divide the
original Hamiltonian (\ref{eq:hamilt2}) into $H=H^{0}+H^{1}$, where $H^{0}$
contains the first two terms of $H$ and $H^{1}$ the last term. Then the
unitary transformation can be proceed order by order, 
\begin{equation}
H^{\prime }=H^{0}+H^{1}+[S,H^{0}]+[S,H^{1}]+{\frac{{1}}{{2}}}%
[S,[S,H^{0}]]+O(\alpha ^{3}).
\end{equation}
The first-order terms in $H^{\prime}$ are 
\begin{eqnarray}
H^{1}+[S,H^{0}] &=&\frac{1}{\sqrt{N}}\sum_{q,k,s}g(k+q,k)(b_{-q}^{\dag
}+b_{q})c_{k+q,s}^{\dag }c_{k,s}  \nonumber \\
&&-\frac{1}{\sqrt{N}}\sum_{q,k,s}g(k+q,k)\delta (k+q,k)(b_{-q}^{\dag
}+b_{q})c_{k+q,s}^{\dag }c_{k,s}  \nonumber \\
&&+\frac{1}{\sqrt{N}}\sum_{q,k,s}\frac{g(k+q,k)}{\omega _{q}}\delta
(k+q,k)(b_{-q}^{\dag }-b_{q})(\epsilon _{k}-\epsilon _{k+q})c_{k+q,s}^{\dag
}c_{k,s}.
\end{eqnarray}
Note that the ground state $|g_{0}\left.\right\rangle$ of $H^{0}$, the
non-interacting system, is a direct product of a filled fermi-sea $|{\rm FS}%
\left.{}\right\rangle$ and a phonon vacuum state $|{\rm ph},0\left.
{}\right\rangle$: 
\begin{equation}
|g_{0}\left. {}\right\rangle =|{\rm FS}\left. {}\right\rangle |{\rm ph},0%
\left.{}\right\rangle .
\end{equation}
Applying the first-order terms on $|g_{0}\left. {}\right\rangle $ we get 
\begin{equation}
(H^{1}+[S,H^{0}])|g_{0}\left. {}\right\rangle =\frac{1}{\sqrt{N}}%
\sum_{q,k,s}gb_{-q}^{\dag }c_{k+q,s}^{\dag }c_{k,s}\left[ 1-\delta
(k+q,k)\left( 1-\frac{\epsilon _{k}-\epsilon _{k+q}}{\omega _{q}}\right) %
\right] |g_{0}\left. {}\right\rangle ,
\end{equation}
since $b_{q}|{\rm ph},0\left. {}\right\rangle =0$. As the band is
half-filled the Fermi energy $\epsilon _{{\rm F}}=0$. Thus $c_{k+q,s}^{\dag
}c_{k,s} |{\rm FS}\left.{}\right\rangle \not=0$ only if $\epsilon _{k+q}\geq
0$ and $\epsilon_{k}\leq 0$. So, we have 
\begin{equation}
(H^{1}+[S,H^{0}])|g_{0}\left. {}\right\rangle =0,
\end{equation}
if we choose 
\begin{equation}
\delta (k+q,k)=1/(1+|\epsilon _{k+q}-\epsilon _{k}|/\omega _{q}).
\end{equation}
This is nothing but making the matrix element of $H^{1}+[S,H^{0}]$ between $%
|g_{0}\left. \right\rangle $ and the lowest-lying excited states vanishing.
Thus the first-order terms, which are not exactly canceled after the
transformation, are related to the higher-lying excited states and should be
irrelevant under renormalization \cite{ra18}. The second-order terms in $%
H^{\prime }$ can be collected as follows: 
\begin{eqnarray}
&&[S,H^{1}]+\frac{1}{2}[S,[S,H^{0}]]  \nonumber \\
&=&\frac{1}{2N}\sum_{q,k,s}\sum_{q^{\prime },k^{\prime }}\frac{%
g(k+q,k)g(k^{\prime }+q^{\prime },k^{\prime })}{\omega _{q}}\delta
(k+q,k)[2-\delta (k^{\prime }+q^{\prime },k^{\prime })]  \nonumber \\
&&\times (b_{-q}^{\dag }-b_{q})(b_{-q^{\prime }}^{\dag }+b_{q^{\prime
}})(c_{k+q,s}^{\dag }c_{k^{\prime },s}\delta _{k,k^{\prime }+q^{\prime
}}-c_{k^{\prime }+q^{\prime },s}^{\dag }c_{k,s}\delta _{k^{\prime },k+q}) 
\nonumber \\
&&+\frac{1}{2N}\sum_{q,k,s}\sum_{q^{\prime },k^{\prime }}\frac{%
g(k+q,k)g(k^{\prime }+q^{\prime },k^{\prime })}{\omega _{q}\omega
_{q^{\prime }}}\delta (k+q,k)\delta (k^{\prime }+q^{\prime },k^{\prime
})(\epsilon _{k+q}-\epsilon _{k})  \nonumber \\
&&\times (b_{-q}^{\dag }-b_{q})(b_{-q^{\prime }}^{\dag }-b_{q^{\prime
}})(c_{k+q,s}^{\dag }c_{k^{\prime },s}\delta _{k,k^{\prime }+q^{\prime
}}-c_{k^{\prime }+q^{\prime },s}^{\dag }c_{k,s}\delta _{k^{\prime },k+q}) 
\nonumber \\
&&-\frac{1}{N}\sum_{q,k,s}\sum_{k^{\prime },s^{\prime }}\frac{%
g(k+q,k)g(k^{\prime }-q,k^{\prime })}{\omega _{q}}\delta (k+q,k)[2-\delta
(k^{\prime }-q,k^{\prime })]c_{k+q,s}^{\dag }c_{k,s}c_{k^{\prime
}-q,s^{\prime }}^{\dag }c_{k^{\prime },s^{\prime }}.
\end{eqnarray}
$\delta _{k^{\prime },k+q}$ is the Kronecker $\delta $ symbol. All terms of
higher order than $\alpha ^{2}$ will be omitted in the following treatment.

Second, for the dimerized state, owing to the spontaneous lattice
distortion, the neighboring atoms move in opposite directions. To take into
account the static phonon-staggered ordering, we make a displacement
transformation to $H^{\prime }$ 
\begin{equation}
H^{\prime \prime }=\exp (R)H^{\prime }\exp (-R).
\end{equation}
Here the generator of the displacement operator $\exp (R)$ is 
\begin{equation}
R=-\sum_{l}(-1)^{l}u_{0}\sqrt{\frac{M\omega _{\pi }}{2}}(b_{l}^{\dag
}-b_{l}).
\end{equation}

If the ground state of $H$ is $|g\left. {}\right\rangle $, then the ground
state of $H^{\prime \prime }$ is $|g^{\prime }\left. {}\right\rangle $: $%
|g\left. {}\right\rangle =e^{-S}e^{-R}|g^{\prime }\left. {}\right\rangle $.
We assume that for $|g^{\prime }\left. {}\right\rangle $ the electrons and
phonons can be decoupled: $|g^{\prime }\left. {}\right\rangle \approx |{\rm e%
}\left. {}\right\rangle |{\rm ph},0\left. {}\right\rangle $, where $|{\rm e}%
\left.{}\right\rangle $ is the ground state for electrons. After averaging $%
H^{\prime \prime }$ over the phonon vacuum state we get an effective
Hamiltonian for the electrons, 
\begin{eqnarray}
H_{{\rm eff}} &=&\left\langle {\rm ph},0|H^{\prime \prime }|{\rm ph}%
,0\right\rangle  \nonumber \\
&=&2NKu_{0}^{2}+\sum_{k,s}E_{0}(k)c_{k,s}^{\dag }c_{k,s}+\sum_{k,s}i\Delta
_{0}(k)c_{k,s}^{\dag }c_{k-\pi ,s}  \nonumber \\
&&-\frac{1}{N}\sum_{q,k,s}\sum_{k^{\prime },s^{\prime }}\frac{%
g(k+q,k)g(k^{\prime }-q,k^{\prime })}{\omega _{q}}\delta (k+q,k)[2-\delta
(k^{\prime }-q,k^{\prime })]c_{k+q,s}^{\dag }c_{k,s}c_{k^{\prime
}-q,s^{\prime }}^{\dag }c_{k^{\prime },s^{\prime }},  \label{heff}
\end{eqnarray}
where 
\begin{eqnarray}
E_{0}(k) &=&\epsilon _{k}-\frac{1}{N}\sum_{k^{\prime }}\frac{g(k^{\prime
},k)g(k,k^{\prime })}{\omega _{k^{\prime }-k}^{2}}\delta (k^{\prime
},k)\delta (k,k^{\prime })(\epsilon _{k}-\epsilon _{k^{\prime }}), \\
\Delta _{0}(k) &=&4\alpha u_{0}\sin k[1-\delta (k-\pi ,k)].
\end{eqnarray}
We find by means of the variational principle that the dimerized lattice
displacement ordering parameter is 
\begin{equation}
u_{0}=-\frac{\alpha }{KN}\sum_{k>0,s}i\sin k[1-\delta (k-\pi
,k)]\left\langle {\rm e}\left| (c_{k,s}^{\dag }c_{k-\pi ,s}-c_{k-\pi
,s}^{\dag}c_{k,s})\right| {\rm e}\right\rangle .
\end{equation}
The last term in $H_{{\rm eff}}$ is a four-fermion interaction. As we are
dealing with a one-dimensional system, how to treat the four-fermion
interaction is a difficult problem.

Note that in the adiabatic limit, where $\omega _{\pi }=0$, one has $\delta
(k^{\prime },k)=0,$ and $H_{{\rm eff}}$ goes back to the adiabatic
mean-field Hamiltonian, 
\begin{equation}
H_{{\rm eff}}(\omega _{\pi }=0)=2NKu_{0}^{2}+\sum_{k,s}\epsilon
_{k}c_{k,s}^{\dag }c_{k,s}+\sum_{k>0,s}i4\alpha u_{0}\sin k(c_{k,s}^{\dag
}c_{k-\pi ,s}-c_{k-\pi ,s}^{\dag }c_{k,s}).  \label{heff1}
\end{equation}
It can be diagonalized exactly by means of the Bogoliubov transformation.
The energy-gap parameter and the phonon-staggered ordering parameter in this
limit can be obtained as 
\begin{equation}
\Delta (k)=4\alpha u_{0}\sin k,
\end{equation}
\begin{equation}
m_{p}=\frac{1}{N}\sum_{l}(-1)^{l}\langle u_{l}\rangle=u_{0}.
\end{equation}
Thus, we have the relation 
\begin{equation}
\Delta_{{\rm ad}}=4\alpha m_{p},
\end{equation}
same as the results of previous works \cite{ra2,ra3,ra16} in the adiabatic
limit.

On the other hand, in the antiadiabatic limit, where $\omega _{\pi
}\rightarrow \infty $, we have $u_{0}=0$, $\delta (k^{\prime },k)=1$, and $%
H_{{\rm eff}}$ becomes 
\begin{equation}
H_{{\rm eff}}(\omega _{\pi }\rightarrow \infty )=\sum_{k,s}\epsilon
_{k}c_{k,s}^{\dag }c_{k,s}-\frac{1}{N}\sum_{q,k,s}\sum_{k^{\prime
},s^{\prime }}\frac{g(k+q,k)g(k^{\prime }-q,k^{\prime })}{\omega _{q}}%
c_{k+q,s}^{\dag }c_{k,s}c_{k^{\prime }-q,s^{\prime }}^{\dag }c_{k^{\prime
},s^{\prime }}.
\end{equation}%
Returning to the real space, this Hamiltonian is 
\begin{eqnarray}
H_{{\rm eff}}(\omega _{\pi } &\rightarrow &\infty
)=-t\sum_{l,s}(c_{l+1,s}^{\dag }c_{l,s}+c_{l,s}^{\dag }c_{l+1,s})  \nonumber
\\
&&\hspace{0.25in}-\frac{\alpha ^{2}}{2K}\sum_{l,s,s^{\prime }}(c_{l,s}^{\dag
}c_{l+1,s}+c_{l+1,s}^{\dag }c_{l,s})(c_{l,s^{\prime }}^{\dag
}c_{l+1,s^{\prime }}+c_{l+1,s^{\prime }}^{\dag }c_{l,s^{\prime }}).
\end{eqnarray}%
For the spinless case, the on-site electron-electron interaction disappears
because of the Pauli principle and this leads the Hamiltonian to be 
\begin{equation}
H_{{\rm eff}}(\omega _{\pi }\rightarrow \infty )=-t\sum_{l}(c_{l}^{\dag
}c_{l+1}+c_{l+1}^{\dag }c_{l})+\frac{\alpha ^{2}}{K}\sum_{l}(c_{l}^{\dag
}c_{l}c_{l+1}^{\dag }c_{l+1}-c_{l}^{\dag }c_{l}),
\end{equation}%
same as the exact effective Hamiltonian obtained by writing the phonon part
of the SSH model as a functional integral and integrating out the phonon
degrees of freedom \cite{ra16}. In this case, the electrons experience an
effective repulsive force between nearest neighbors, and for $\alpha
^{2}/2Kt=1$ the system undergoes a transition to a charge-density-wave
state. This also is equivalent to the antiferromagnetic $XXZ$ spin chain
model (via a Jordan-Wigner transformation) with the phase transition taking
place at the isotropic point \cite{ra24}. Thus our effective Hamiltonian
works well in both the $\omega _{\pi }=0$ and the $\omega _{\pi }\rightarrow
\infty $ limits.

Now the total Hamiltonian can be written as $\tilde{H}=\tilde{H}_{0}+\tilde{%
H
}_{1}$, where $\tilde{H}_{1}$ includes the terms which are zero after
being averaged over the phonon vacuum state, and 
\begin{equation}
\tilde{H}_{0}=\sum_{q}\omega _{q}\left( b_{q}^{\dag }b_{q}+\frac{1}{2}%
\right) +H_{{\rm eff}}.
\end{equation}
Through above two unitary transformations, the Hamiltonian is divided into
two parts: $\tilde{H}_{0}$ and $\tilde{H}_{1}$. We believe that the $\tilde{H%
}_{0}$ contains the main physics of the coupling system. This is supported
by: (a) $H_{{\rm eff}}$ [Eq. (\ref{heff})] works well for the both limits $%
\omega _{\pi}=0$ and $\omega _{\pi}\rightarrow \infty $, even if the
electron-phonon coupling is strong; (b) the effect of the first order terms
in $H^{\prime\prime}$ is eliminated in the lowest order of perturbation by
introducing a function $\delta(k^{\prime},k)$; (c) the results for the
general $\omega _{\pi}$ in the spinless case are compared quite well \cite%
{ra18} with those of the Monte Carlo simulations \cite{ra16}.

\section{Analytical Approach}

Note that the four-fermion interaction term in Eq. (\ref{heff}) goes to zero
as $\omega _{\pi }\rightarrow 0$ [see Eq. (\ref{heff1})]. Therefore, in the
case of small $\omega_{\pi }$, this term can be treated as a perturbation,
and the unperturbed Hamiltonian is 
\begin{equation}
H_{{\rm eff}}^{0}=2NKu_{0}^{2}+\sum_{k,s}E_{0}(k)c_{k,s}^{\dag
}c_{k,s}+\sum_{k,s}i\Delta _{0}(k)c_{k,s}^{\dag }c_{k-\pi ,s}.
\end{equation}
The four-fermion term can be re-written as 
\begin{eqnarray}
H_{{\rm eff}}^{\prime } &=&-\frac{1}{N}\sum_{q,k,s}\sum_{k^{\prime
},s^{\prime }}\frac{g(k+q,k)g(k^{\prime }-q,k^{\prime })}{\omega _{q}}\delta
(k+q,k)[2-\delta (k^{\prime }-q,k^{\prime })]  \nonumber \\
&&\times (c_{k+q,s}^{\dag }c_{k,s}-c_{k+q-\pi ,s}^{\dag }c_{k-\pi
,s})(c_{k^{\prime }-q,s^{\prime }}^{\dag }c_{k^{\prime },s^{\prime
}}-c_{k^{\prime }-q-\pi ,s^{\prime }}^{\dag }c_{k^{\prime }-\pi ,s^{\prime
}})  \nonumber \\
&&+\frac{1}{N}\sum_{q,k,s}\sum_{k^{\prime },s^{\prime }}\frac{g(k+q,k-\pi
)g(k^{\prime }-q,k^{\prime }-\pi )}{\omega _{q-\pi }}\delta (k+q,k-\pi
)[2-\delta (k^{\prime }-q,k^{\prime }-\pi )]  \nonumber \\
&&\times (c_{k+q,s}^{\dag }c_{k-\pi ,s}c_{k^{\prime }-q-\pi ,s^{\prime
}}^{\dag }c_{k^{\prime },s^{\prime }}+c_{k+q-\pi ,s}^{\dag
}c_{k,s}c_{k^{\prime }-q,s^{\prime }}^{\dag }c_{k^{\prime }-\pi ,s^{\prime
}})  \nonumber \\
&&-\frac{1}{N}\sum_{q,k,s}\sum_{k^{\prime },s^{\prime }}\frac{g(k+q,k-\pi
)g(k^{\prime }-q,k^{\prime }-\pi )}{\omega _{q-\pi }}\delta (k+q,k-\pi
)[2-\delta (k^{\prime }-q,k^{\prime }-\pi )]  \nonumber \\
&&\times (c_{k+q-\pi ,s}^{\dag }c_{k,s}c_{k^{\prime }-q-\pi ,s^{\prime
}}^{\dag }c_{k^{\prime },s^{\prime }}+c_{k+q,s}^{\dag }c_{k-\pi
,s}c_{k^{\prime }-q,s^{\prime }}^{\dag }c_{k^{\prime }-\pi ,s^{\prime }}).
\label{eq:Hamil}
\end{eqnarray}
In these terms we have the constraints 
\[
k>0,\ \ \ \ k^{\prime }>0,\ \ \ \ k+q>0,\ \ \ \ k^{\prime }-q>0. 
\]
We can distinguish between different physical processes. The first term in
Eq. (\ref{eq:Hamil}) is the forward scattering one, the second the backward
one, and the last the Umklapp one. We use the Green's function method to
implement the perturbation treatment and it is more convenient to work
within a two-component representation, 
\begin{equation}
\Psi _{k,s}=\left( {\ 
\begin{array}{l}
c_{k,s} \\ 
c_{k-\pi ,s}%
\end{array}
}\right) ,\ \ \ \ \ \ \ \ \ k>0.
\end{equation}
Thus we have 
\begin{equation}
\left\{ {\ 
\begin{array}{l}
\Psi _{k,s}^{\dag }\sigma _{z}\Psi _{k,s}=c_{k,s}^{\dag }c_{k,s}-c_{k-\pi
,s}^{\dag }c_{k-\pi ,s} \\ 
\Psi _{k,s}^{\dag }\sigma _{x}\Psi _{k,s}=c_{k,s}^{\dag }c_{k-\pi
,s}+c_{k-\pi ,s}^{\dag }c_{k,s} \\ 
\Psi _{k,s}^{\dag }i\sigma _{y}\Psi _{k,s}=c_{k,s}^{\dag }c_{k-\pi
,s}-c_{k-\pi ,s}^{\dag }c_{k,s}%
\end{array}
}\right. ,
\end{equation}
and the Hamiltonian becomes 
\begin{equation}
H_{{\rm eff}}^{0}=2NKu_{0}^{2}+\sum_{k,s}E_{0}(k)\Psi_{k,s}^{\dag}\sigma
_{z}\Psi_{k,s}-\sum_{k,s}\Delta _{0}(k)\Psi _{k,s}^{\dag }\sigma _{y}\Psi
_{k,s},
\end{equation}
and 
\begin{eqnarray}
H_{{\rm eff}}^{^{\prime }} &=&-\frac{1}{N}\sum_{q,k,s}\sum_{k^{\prime
},s^{\prime }}\frac{g(k+q,k)g(k^{\prime }-q,k^{\prime })}{\omega _{q}}\delta
(k+q,k)[2-\delta (k^{\prime }-q,k^{\prime })]  \nonumber \\
&&\times \Psi _{k+q,s}^{\dag }\sigma _{z}\Psi _{k,s}\Psi _{k^{\prime
}-q,s^{\prime }}^{\dag }\sigma _{z}\Psi _{k^{\prime },s^{\prime }}  \nonumber
\\
&&+\frac{1}{2N}\sum_{q,k,s}\sum_{k^{\prime },s^{\prime }}\frac{g(k+q,k-\pi
)g(k^{\prime }-q,k^{\prime }-\pi )}{\omega _{q-\pi }}\delta (k+q,k-\pi
)[2-\delta (k^{\prime }-q,k^{\prime }-\pi )]  \nonumber \\
&&\times (\Psi _{k+q,s}^{\dag }\sigma _{x}\Psi _{k,s}\Psi _{k^{\prime
}-q,s^{\prime }}^{\dag }\sigma _{x}\Psi _{k^{\prime },s^{\prime }}-\Psi
_{k+q,s}^{\dag }i\sigma _{y}\Psi _{k,s}\Psi _{k^{\prime }-q,s^{\prime
}}^{\dag }i\sigma _{y}\Psi _{k^{\prime },s^{\prime }})  \nonumber \\
&&-\frac{1}{2N}\sum_{q,k,s}\sum_{k^{\prime },s^{\prime }}\frac{g(k+q,k-\pi
)g(k^{\prime }-q,k^{\prime }-\pi )}{\omega _{q-\pi }}\delta (k+q,k-\pi
)[2-\delta (k^{\prime }-q,k^{\prime }-\pi )]  \nonumber \\
&&\times (\Psi _{k+q,s}^{\dag }\sigma _{x}\Psi _{k,s}\Psi _{k^{\prime
}-q,s^{\prime }}^{\dag }\sigma _{x}\Psi _{k^{\prime },s^{\prime }}+\Psi
_{k+q,s}^{\dag }i\sigma _{y}\Psi _{k,s}\Psi _{k^{\prime }-q,s^{\prime
}}^{\dag }i\sigma _{y}\Psi _{k^{\prime },s^{\prime }}).
\end{eqnarray}
$\sigma _{\beta }(\beta =x,y,z)$ is the Pauli matrix. The matrix Green's
function is defined as (the temperature Green's function is used and at the
end let $T\rightarrow 0$) 
\begin{eqnarray}
G_{s,s^{\prime }}(k,\tau ) &=&-\langle T_{\tau }\Psi _{k,s}(\tau )\Psi
_{k,s^{\prime }}^{\dag }(0)\rangle  \nonumber \\
&=&\frac{1}{\beta }\sum_{n}\exp (-i\omega _{n}\tau )G_{s,s^{\prime
}}(k,\omega _{n}).
\end{eqnarray}
The Dyson equation is 
\begin{equation}
G_{s,s^{\prime }}(k,\omega _{n})=G_{s,s^{\prime }}^{0}(k,\omega
_{n})+G_{s,\alpha }^{0}(k,\omega _{n})\Sigma _{\alpha ,\beta }^{\ast
}(k,\omega _{n})G_{\beta ,s^{\prime }}(k,\omega _{n}),
\end{equation}
where 
\begin{equation}
G_{s,s^{\prime }}^{0}(k,\omega _{n})=\delta _{s,s^{\prime }}\left\{ i\omega
_{n}-E_{0}(k)\sigma _{z}+\Delta _{0}(k)\sigma _{y}\right\} ^{-1},
\end{equation}
is the unperturbed Green's function. The self-energy $\Sigma _{\alpha ,\beta
}^{\ast }(k,\omega _{n})$ can be calculated by the perturbation theory, 
\begin{eqnarray}
\Sigma _{\alpha ,\beta }^{\ast }(k,\omega _{n}) &=&\frac{T}{N}%
\sum_{k^{\prime }>0}\sum_{m}\frac{g(k,k^{\prime})g(k^{\prime},k)}{\omega
_{k^{\prime}-k}}\delta (k^{\prime},k)[2-\delta (k,k^{\prime })]\{G_{\alpha
,\beta }^{0}(k^{\prime },\omega_{m})+\sigma _{z}G_{\alpha ,\beta
}^{0}(k^{\prime },\omega _{m})\sigma _{z}\}  \nonumber \\
&&+\frac{T}{N}\sum_{k^{\prime }>0}\sum_{m}\frac{g(k,k^{\prime}-\pi)g(k^{%
\prime},k-\pi)}{\omega _{k^{\prime}-k-\pi}}\delta(k^{\prime },k-\pi
)[2-\delta (k,k^{\prime }-\pi)]  \nonumber \\
&&\times \{i\sigma _{y}G_{\alpha ,\beta }^{0}(k^{\prime },\omega
_{m})i\sigma _{y}-\sigma _{x}G_{\alpha ,\beta }^{0}(k^{\prime },\omega
_{m})\sigma _{x}\}  \nonumber \\
&&-\frac{T}{N}\sum_{k^{\prime }}\sum_{m}\frac{g(k,k-\pi)g(k^{\prime},k^{%
\prime}-\pi)}{\omega _{\pi}}[\delta(k,k-\pi )+\delta (k^{\prime },k^{\prime
}-\pi )-\delta (k,k-\pi )\delta(k^{\prime },k^{\prime }-\pi )]  \nonumber \\
&&\times \left\{ T_{r}[i\sigma _{y}G_{\alpha ,\beta }^{0}(k^{\prime },\omega
_{m})]i\sigma _{y}+T_{r}[\sigma _{x}G_{\alpha ,\beta }^{0}(k^{\prime
},\omega _{m})]\sigma _{x}\right\} .  \label{eq:sener}
\end{eqnarray}
In the theoretical analysis we have taken into account the fact that only
the Umklapp scattering terms affect the gap, and the forward and backward
scattering terms contribute nothing to the ''charge'' gap \cite{ra25,ra26}.
From Eq. (\ref{eq:sener}) one can get that $\Sigma _{\alpha ,\beta }^{\ast
}(k,\omega _{n})$ is irrelative to $\omega _{n}$, therefore the spectrum
structure of $G_{s,s^{\prime }}(k,\omega _{n})$ should be 
\begin{equation}
G_{s,s^{\prime }}(k,\omega _{n})=\delta _{s,s^{\prime }}\left\{ i\omega
_{n}-E(k)\sigma _{z}+\Delta (k)\sigma _{y}\right\} ^{-1}.
\end{equation}
From $G_{s,s^{\prime}}(k,\omega _{n})$ the elementary excitation spectrum in
the gapped state can be derived 
\begin{equation}
W(k)=\sqrt{E^{2}(k)+\Delta ^{2}(k)}.
\end{equation}
The renormalized band function and the gap function are 
\begin{eqnarray}
E(k) &=&E_{0}(k)-\frac{1}{N}\sum_{k^{\prime }}\frac{2\alpha ^{2}}{K}\cos
^{2}\left( \frac{k^{\prime }+k}{2}\right) \delta (k^{\prime },k)[2-\delta
(k^{\prime },k)]\frac{E_{0}(k^{\prime })}{\sqrt{E_{0}^{2}(k^{\prime
})+\Delta _{0}^{2}(k^{\prime })}}, \\
\Delta (k) &=&4\alpha u_{0}\sin k[c-d\delta (k-\pi ,k)].  \label{delta}
\end{eqnarray}
Here 
\begin{eqnarray}
c &=&1+\frac{1}{N}\sum_{k>0,s}\frac{\alpha ^{2}}{K}\delta (k-\pi ,k)\sin k%
\frac{\Delta _{0}(k)}{2\alpha u_{0}\sqrt{E_{0}^{2}(k)+\Delta _{0}^{2}(k)}},
\\
d &=&1-\frac{1}{N}\sum_{k>0,s}\frac{\alpha ^{2}}{K}[1-\delta (k-\pi ,k)]\sin
k\frac{\Delta _{0}(k)}{2\alpha u_{0}\sqrt{E_{0}^{2}(k)+\Delta _{0}^{2}(k)}}.
\end{eqnarray}
The equation to determine $u_{0}$ is 
\begin{equation}
1=\frac{1}{N}\sum_{k>0,s}\frac{\alpha ^{2}}{K}[1-\delta (k-\pi ,k)]\sin k%
\frac{\Delta (k)}{\alpha u_{0}W(k)}.
\end{equation}
These are basic equations in our theory.

The phonon-staggered ordering parameter 
\begin{eqnarray}
m_{p} & = & \frac{1}{N}\sum_{l}(-1)^{l}\langle u_{l}\rangle  \nonumber \\
& = & \frac{1}{N}\sum_{k>0,s}\frac{\alpha}{K}\sin k\frac{\Delta(k)}{W(k)}.
\label{mppp}
\end{eqnarray}

If $\omega _{\pi }=0$ we have $\delta (k^{\prime },k)=0$ and $c=1$, Eq. (\ref%
{delta}) becomes the same as that in the adiabatic theory $\Delta_{{\rm ad}%
}= \Delta(\pi /2)=4\alpha u_{0}$. Comparing Eq. (\ref{delta}) with the
energy gap in the adiabatic case, $\Delta =4\alpha u_{0}$, we have the gap
in the nonadiabatic case, $\Delta =\Delta (\pi /2)=4\alpha u_{0}[c-d]$. This
is the true gap in the excitation spectrum.

Fig. 1 shows the elementary excitation spectrum $W(k)$ as function of the
wave vector in the case of $\alpha ^{2}/Kt=0.336$ for different phonon
frequencies $\omega _{\pi }/t=0.001$, $0.01$, $0.1$, and $0.2$. The
adiabatic limit result is also shown in dash-dot-dot line for comparison.
Near the Fermi surface, the change of phonon frequency strongly affects the
elementary excitation spectrum. The dimerization gap decreases as the phonon
frequency increases and changes very sensitively to the change of phonon
frequency, especially when phonon frequency is small. The dimerization gap
is much more reduced by the quantum lattice fluctuations than other order
parameters, for instance, the phonon-staggered ordering parameter. The
reason for this will be discussed later. In the mean field approximation the
Peierls distortion opens a gap $2\Delta _{{\rm MF}}$ and $\Delta _{{\rm MF}%
}=4\alpha m_{p}$. Our results indicate that this relation holds only in the
adiabatic limit. The effect of the quantum lattice fluctuations on the
dimerization gap can be seen clearly also in the mean electron occupation
number 
\begin{equation}
\sum_{s}\langle c_{k,s}^{\dagger }c_{k,s}\rangle=1-\frac{E(k)}{W(k)}\,,
\end{equation}%
Fig. 2 shows the mean electron occupation number as function of the wave
vector in the cases of $\alpha ^{2}/Kt=0.2$ with $\omega _{\pi }/t=0$, and $%
\alpha ^{2}/Kt=0.4$ with $\omega _{\pi }/t=0$ and $0.2$. The more the phonon
frequency or the electron-phonon coupling increases the more the Fermi
surface is smeared. The strongest effect of the quantum phonon fluctuation
at the Fermi surface leads to the strongest modulation of the dimerization
gap. When $\alpha =0$, the system becomes one for free electrons and there
is no long-range order, and consequently, as shown in dash-dot-dot line, the
system has a clear Fermi surface at $k=\pi /2$. When $\alpha >0$, the
effective electron-electron interactions induced by the electron-phonon
coupling correlate the electrons and, thereby, the excitation spectrum of
the system becomes one of quasi-particle excitations superseding the single
electron excitations in $\alpha =0$ case.

The density-of-states of fermions $\rho (\omega )$ is 
\begin{eqnarray}
\rho (\omega ) &=&\frac{1}{N}\sum_{k,s}\delta \left( \omega -\sqrt{%
E^{2}(k)+\Delta ^{2}(k)}\right)  \nonumber \\
&=&\frac{1}{\pi }\left( \left. \frac{d}{dk}\sqrt{E^{2}(k)+\Delta ^{2}(k)}%
\right| _{k=f(\omega )}\right) ^{-1},
\end{eqnarray}%
where, $k=f(\omega )$ is the inverse function of $\omega =\sqrt{%
E^{2}(k)+\Delta ^{2}(k)}$. Fig. 3 shows the calculated DOS of fermions for
electron-phonon coupling $\alpha ^{2}/Kt=0.336$ with $\omega _{\pi
}/t=0.001,0.01$, and $0.04$, respectively. One can see that a nonzero DOS
stars from the gap edge and, for small values of $\omega _{\pi }$, there is
a peak above the gap edge with a significant tail between it and the true
gap edge. As $\omega _{\pi }$ increases, the peak height of the DOS
decreases, and the peak broadens and moves to lower photon energy which
implies that the dimerization gap becomes narrower and the Fermi surface
becomes more smear. The inverse-square-root singularity at the gap edge in
the adiabatic case \cite{ra3,ra15} disappears. For comparison, the adiabatic 
$(\omega _{\pi }=0)$ result is also shown in the dash-dot line which has an
inverse-square-root singularity at gap edge. A previous analytic treatment
on the modification of DOS due to lattice fluctuation showed the move of the
DOS peak, but there existed still the singularity \cite{ra7}. Our results
consist with the spectrum measurements in both the peak position and the
line shape of DOS.

The mean phonon occupation number (MPON) is 
\begin{equation}
\frac{1}{N}\sum_{q}\langle b_{q}^{\dagger }b_{q}\rangle= -\frac{2K}{%
\omega_{\pi}} u^{2}_{0} +\frac{1}{N}\sum_{k>0,s}\frac{4\alpha u_{0}}{%
\omega_{\pi}}\sin k \frac{\Delta(k)}{W(k)} +\frac{1}{N^{2}}%
\sum_{k>0,s}\sum_{k^{\prime}>0}^{\pi/2}\frac{4\alpha^{2}}{K} \frac{%
\delta^{2}(k,k^{\prime})}{\omega_{k^{\prime}-k}}\cos^{2}\frac{k+k^{\prime}}{2%
}.
\end{equation}
Fig. 4 shows the dependences of MPON on the phonon frequency for $%
\alpha^{2}/Kt=0.336, 0.50$ and $0.81$, respectively. MPON decreases as the
electron-phonon coupling decreases or the phonon frequency increases. For
small $\omega _{\pi }$, MPON changes very sensitively to the phonon
frequency, and when $\omega _{\pi }=0$, MPON is divergent. Therefore, in
point of the MPON, adiabatic approach is corresponding to the divergence of
MPON, while for realistic materials, MPON is finite or even small.

\section{Optical Absorption Coefficient}

The optical absorption has remained an important source of information about
the excitation spectrum of (CH)$_{x}$ \cite{ra27}, and the measurements of
it's optical properties have provided much of the experimental basis for
theoretical studies \cite{ra28}. In the pristine sample the absorption curve
peaks at about 2 eV and the peak is followed by a broad tail extending into
the gap. To appreciate the nonadiabatic effects on the optical excitation
spectrum of (CH)$_{x}$, we further calculate the optical absorption
coefficient under nonadiabatic circumstance by adopting the SSH model. The
optical absorption coefficient $\alpha (\omega )$ can be expressed by the
retarded Green's function \cite{ra29} as follows: 
\begin{equation}
\alpha (\omega )=-\frac{2}{\pi \omega }{\rm Im}K^{R}(\omega ),
\end{equation}%
where $K^{R}$ is defined as 
\begin{equation}
K^{R}(\omega )=-i\int_{-\infty }^{0}e^{-i\omega t}dt\langle
g|[j(0)j(t)-j(t)j(0)]|g\rangle .
\end{equation}%
Here $j$ is the current operator \cite{ra30}, 
\begin{eqnarray}
j &=&iev_{F}\sum_{l,s}(c_{l,s}^{\dag }c_{l+1,s}-c_{l+1,s}^{\dag }c_{l,s}) 
\nonumber \\
&=&J\sum_{k>0,s}\sin k\Psi _{k,s}^{\dag }\sigma _{z}\Psi _{k,s},
\end{eqnarray}%
$v_{F}=2at$,$J=2ev_{F}$, and $j(t)=\exp (iHt)j\exp (-iHt)$ is the form of $j$
in the Heisenberg representation. Because the averaging of $\tilde{H_{1}}$
over the phonon vacuum state is zero, in the ground state at zero
temperature $\tilde{H_{1}}$ can be neglected. Then, by using the
approximately decoupling $|g^{\prime }\rangle \approx |g_{0}^{\prime
}\rangle $, the ground state of $\tilde{H_{0}}$, and $\tilde{H}\approx 
\tilde{H}_{0}$ in the calculation 
\begin{eqnarray}
\langle g|j(0)j(t)|g\rangle &=&\langle g^{\prime }|[e^{(S+R)}je^{-(S+R)}]e^{i%
\tilde{H}t}[e^{(S+R)}je^{-(S+R)}]e^{-i\tilde{H}t}|g^{\prime }\rangle 
\nonumber \\
&\approx &\langle g_{0}^{\prime }|[e^{S}je^{-S}]e^{i\tilde{H}%
_{0}t}[e^{S}je^{-S}]e^{-i\tilde{H}_{0}t}|g_{0}^{\prime }\rangle ,
\end{eqnarray}%
we can get 
\begin{eqnarray}
K^{R}(\omega ) &=&\frac{J^{2}}{N}\sum_{k>0,s}\left( \frac{1}{\omega
-2W(k)+i0^{+}}-\frac{1}{\omega +2W(k)-i0^{+}}\right)  \nonumber \\
&&\times \left[ \sin ^{2}k-\frac{2}{N}\sum_{k^{\prime }}\frac{g(k^{\prime
},k)g(k,k^{\prime })}{\omega _{k-k^{\prime }}^{2}}\delta ^{2}(k^{\prime
},k)\sin k(\sin k-\sin k^{\prime })\right] \frac{\Delta ^{2}(k)}{W^{2}(k)} 
\nonumber \\
&&+\frac{J^{2}}{N^{2}}\sum_{k>0,k^{\prime },s}\left( \frac{1}{\omega -\omega
_{k-k^{\prime }}-W(k)-W(k^{\prime })+i0^{+}}-\frac{1}{\omega +\omega
_{k-k^{\prime }}+W(k)+W(k^{\prime })-i0^{+}}\right)  \nonumber \\
&&\times \frac{g(k^{\prime },k)g(k,k^{\prime })}{\omega _{k-k^{\prime }}^{2}}%
\delta ^{2}(k^{\prime },k)(\sin k-\sin k^{\prime })^{2}\left[ 1-{\rm sign}%
k^{\prime }\frac{E(k)E(k^{\prime })+\Delta (k)\Delta (k^{\prime })}{%
W(k)W(k^{\prime })}\right] .
\end{eqnarray}%
Thus, the optical absorption coefficient 
\begin{eqnarray}
\alpha (\omega ) &=&\frac{2J^{2}}{\omega N}\sum_{k>0,s}\delta \lbrack \omega
-2W(k)]\frac{\Delta ^{2}(k)}{W^{2}(k)}  \nonumber \\
&&\times \left[ \sin ^{2}k-\frac{2}{N}\sum_{k^{\prime }}\frac{4\alpha ^{2}}{K%
}\cos ^{3}\frac{k+k^{\prime }}{2}{\rm sign}\left( \sin \frac{k-k^{\prime }}{2%
}\right) \frac{\omega _{\pi }\sin k}{(\omega _{\pi }+4t|\sin \frac{%
k-k^{\prime }}{2}|)^{2}}\right]  \nonumber \\
&&+\frac{4J^{2}}{\omega N^{2}}\sum_{k>0,s}\sum_{k^{\prime }}\delta \lbrack
\omega -\omega _{k-k^{\prime }}-W(k)-W(k^{\prime })]\left[ 1-{\rm sign}%
k^{\prime }\frac{E(k)E(k^{\prime })+\Delta (k)\Delta (k^{\prime })}{%
W(k)W(k^{\prime })}\right]  \nonumber \\
&&\times \frac{8\alpha ^{2}}{K}\cos ^{4}\frac{k+k^{\prime }}{2}\left| \sin 
\frac{k-k^{\prime }}{2}\right| \frac{\omega _{\pi }}{(\omega _{\pi }+4t|\sin 
\frac{k-k^{\prime }}{2}|)^{2}},  \label{alf}
\end{eqnarray}%
and the $\omega -$integrated spectral-weight function 
\begin{equation}
S(\omega )=\int_{0}^{\omega }\alpha (\omega ^{\prime })d\omega ^{\prime }.
\end{equation}

In Fig. 5, we compare our calculation (solid line) with the observed optical
absorption (solid circles) \cite{ra10} of {\it trans}-(CH)$_{x}$ polymer
chain. The optical absorption coefficient $\alpha (\omega )$ is normalized
to its peak value $\alpha (\omega _{{\rm p}})$ and the photon energy $\omega$
is in unit of its peak value $\omega _{{\rm p}}$ for convenience to
comparison. Here we use the input parameters $\alpha =4.1$ eV/\AA, $K=21$
eV/\AA$^{2}$, $t=2.5$ eV, same as those of Su, Schrieffer and Heeger \cite%
{ra1}, and $\omega _{\pi }/t=0.0652$. One can see that the agreement between
the experiment result and our calculation is quite good. The relationships
between the changes of the optical absorption shape as well as its peak
position and the different phonon frequencies can be seen clearly in Fig. 6.
The parameter values used are: $\alpha ^{2}/Kt=0.35$, $K=21$ eV/\AA$^{2}$,
and $t=2.5$ eV with $\omega _{\pi }/t=0.002$ (solid line) and $0.04$ (dash
line). The calculated true dimerization gaps relating to these two sets of
parameter values are marked by the arrows. The solid circles denote the
experimental results obtained from analysis of the inelastic
electron-scattering data \cite{ra3}. One can see that the spectrum broadens
but the peak height decreases and shifts to lower photon energy as $\omega
_{\pi }$ increases. The inverse-square-root singularity at the gap edge in
the adiabatic case \cite{ra3,ra29,ra30} and in the other methods \cite{ra7}
disappears because of the nonadiabatic effect. For finite $\omega _{\pi }$,
there exists a significant tail below the peak. This figure shows again the
excellent consistency of the optical absorption coefficient between our
calculation and the experiment result in not only the peak position but also
the line shape. It should be pointed out that in our calculation of the
optical absorption coefficient, the choice of the parameter value set for
the solid line requires $\alpha =4.28$ eV/\AA beening a little bit larger
than $\alpha =4.1$ eV/\AA used in Ref. \cite{ra1}, which implies that in our
theory, due to the nonadiabatic effect, the larger electron-phonon coupling
is required than that in adiabatic approach to obtain the same optical
excitation energy. In addition, in experiments the measurement of optical
absorption spectrum could be affected by various factors, such as the
impurity in samples, the finite measurement temperature, and the finite
measurement resolution. These may make the measured absorption spectrum to
be broadened and lead to the slight discrepancy between the experiment
result and our calculation.

Taking the excellent agreement of our results with the experiment
observations on the optical absorption coefficient as the check and
verification of the effectiveness of our analytical method, we further study
the difference between the dimerization gap and the optical excitation
response energy of the {\it trans}-(CH)$_{x}$ polymer chain within the SSH
model. The experiments of many quasi-one-dimensional materials showed that
the energy gap deduced from the absorption edge is smaller than the
activation energy of the dc conductivity \cite{ra31}, which, as was pointed
out \cite{ra11,ra12}, can not be explained in methods assuming a static
lattice distortion. A Monte Carlo calculation also suggested that the
optical gap can be much larger than the dimerization gap \cite{ra32}. Our
calculated peak position of optical absorption coefficient $\omega _{{\rm p}%
} $ and the true dimerization gap $2\Delta =2\Delta (\pi /2)$ as functions
of phonon frequency $\omega _{\pi }$ are illustrated in Fig. 7. One can see
that for the finite phonon frequency, the peak position of optical
absorption coefficient is not directly corresponding to the true
dimerization gap. The dimerization energy gap is smaller than the optical
gap. It is the most notable that there is a discontinuous drop in the
dimerization gap once the phonon frequency changes, no matter how small it
is, from zero to finite, though at the adiabatic limit the dimerization gap $%
2\Delta (\pi /2)=\omega _{{\rm p}}$. After the drop the dimerization gap and
the peak position of the optical absorption coefficient decrease as the
phonon frequency increases. The dimerization gap is much more reduced by the
quantum lattice fluctuations than the optical gap is. The dimerization gap
is the value of Eq. (\ref{delta}) at the Fermi point $k=\pi /2$, where the
quantum lattice fluctuations have the strongest effect, while the optical
absorption coefficient is the integral [see Eq. (\ref{alf})] over the all
Brillouin zone and the effect of the fluctuations is gentled. We note that
in our theory, in mathematical viewpoint, the difference of the dimerization
gap between the $\omega _{\pi }=0$ and $\omega _{\pi }>0$ cases mainly comes
from the functional form of the gap [see Eq. (\ref{delta})]. Comparing it
with that in the adiabatic limit, one can see that the subgap states come
from the quantum lattice fluctuations, i.e., the second term in the square
bracket of Eq. (\ref{delta}). Substituting $\omega _{\pi }=0$ into Eq. (\ref%
{alf}), we obtain the optical absorption coefficient in the adiabatic limit 
\begin{equation}
\alpha (\omega )=\frac{8J^{2}}{\pi \omega ^{3}}(4\alpha u_{0})^{2}\sin
^{4}k\left. \left[ \frac{d\omega }{dk}\right] ^{-1}\right| _{k=\phi (\omega
)},  \label{adalp}
\end{equation}%
where, $k=\phi (\omega )$ is the inverse function of $\omega =2\sqrt{%
\epsilon _{k}^{2}+(4\alpha u_{0}\sin k)^{2}}$. Obviously, $k=\pi /2$ is a
singular point of Eq. (\ref{adalp}), and, at this point, $\omega _{{\rm p}%
}=2(4\alpha u_{0})=2\Delta $. Thus we get the same relation as in the
adiabatic approach methods assuming a static lattice distortion, but our
result indicates that this relation holds only in the adiabatic limit. In
nonadiabatic case, in our theory, the logarithmic singularity in the
integration of Eq. (\ref{alf}) in the adiabatic case is removed by the
factor $1-\delta (k-\pi ,k)$ as long as $\omega _{\pi}$ is finite. The
effects of quantum lattice fluctuations on the dimerization gap and on the
optical gap are essentially different, especially when $\omega _{\pi }$ is
small. Though some works ascribed the difference between the dimerization
gap and the optical excitation response energy to the electron-electron
interactions \cite{ra10}, our calculation reveals that the effects of the
quantum phonon can lead the dimerization gap to be much smaller than the
optical gap in electron-phonon coupling systems. Contrary to the statement
mentioned in the beginning that the quantum lattice fluctuations cause a
slightly small reduction of the dimerization order parameter \cite%
{ra15,ra16,ra17}, our analysis shows that the dimerization gap being much
more reduced behaves very different from other order parameters.

\section{Conclusions}

We have studied the dimerization gap and the optical absorption coefficient
of the Su-Schrieffer-Heeger model by means of developing an analytical
nonadiabatic approach. By investigating quantitatively the effects of
quantum phonon fluctuations on the gap order and the optical responses in
this system, we have shown that the dimerization gap is much more reduced by
the quantum lattice fluctuations than the optical absorption coefficient is.
The calculated optical absorption coefficient and the density of states do
not have the inverse-square-root singularity, but have a peak above the gap
edge and there exists a significant tail below the peak. The peak of optical
absorption spectrum is not directly corresponding to the dimerized gap. Our
results of the optical absorption coefficient agree well with those of the
experiments in both the shape and the peak position of the optical
absorption spectrum. When the phonon frequency $\omega _{\pi}>0$, there
still exists a static dimerization of the lattice but the quantum lattice
fluctuations (the zero-point lattice motion) play a very important role.
Taking the SSH model as a typical example for electron-phonon coupling
systems in this study, we show that the strongest effects of the quantum
lattice fluctuations at the Fermi surface lead the dimerization gap to be
much smaller than the optical gap in electron-phonon coupling systems. The
difference between the true dimerization gap and the peak position of
optical absorption spectrum is obtained quantitatively.

This work was supported by NSF of China.

\newpage {\bf {\Large Figure Caption}}\newline

{\bf Fig. 1} The fermionic excitation spectrum $W(k)$ as function of the
wave vector in the case of $\alpha^{2}/Kt=0.336$ for different phonon
frequencies $\omega_{\pi}/t=0.001$, $0.01$, $0.1$, and $0.2$. The adiabatic
limit result is also shown in dash-dot-dot line for comparison.\newline

{\bf Fig. 2} The mean electron occupation number as function of the wave
vector in the cases of $\alpha^{2}/Kt=0.2$ with $\omega_{\pi}/t=0$, and $%
\alpha^{2}/Kt=0.4$ with $\omega_{\pi}/t=0$ and $0.2$. The result for $%
\alpha=0$ is shown in dash-dot-dot line.\newline

{\bf Fig. 3} The calculated DOS of fermions for electron-phonon coupling $%
\alpha^{2}/Kt=0.336$ with $\omega_{\pi}/t=0.001, 0.01$, and $0.04$,
respectively. For comparison, the adiabatic $(\omega_{\pi}=0)$ result is
also shown in the dash-dot line.\newline

{\bf Fig. 4} The dependences of MPON on the phonon frequency for $%
\alpha^{2}/Kt=0.336, 0.50$ and $0.81$, respectively.\newline

{\bf Fig. 5} The comparison of our calculation (solid line) with the
observed optical absorption (solid circles) of {\it trans}-(CH)$_{x}$
polymer chain. The optical absorption coefficient $\alpha (\omega )$ is
normalized to its peak value $\alpha (\omega _{{\rm p}})$ and the photon
energy $\omega $ is in unit of its peak value $\omega _{{\rm p}})$ for
convenience to comparison. The input parameters are $\alpha =4.1$ eV/\AA, $%
K=21$ eV/\AA$^{2}$, $t=2.5$ eV, and $\omega _{\pi }/t=0.0652$.\newline

{\bf Fig. 6} The relationships between the changes of the optical absorption
shape as well as its peak position and the different phonon frequencies. The
parameter values used are: $\alpha^{2}/Kt=0.35$, $K=21$ eV/\AA$^{2}$, and $%
t=2.5$ eV with $\omega_{\pi}/t=0.002$ (solid line) and $0.04$ (dash line).
The calculated true dimerization gaps relating to these two sets of
parameter values are marked by the arrows. The solid circles denote the
experimental results obtained from analysis of the inelastic
electron-scattering data.\newline

{\bf Fig. 7} The calculated peak position of optical absorption coefficient $%
\omega_{{\rm p}}$ and the true dimerization gap $2\Delta=2\Delta (\pi /2)$
as functions of phonon frequency $\omega_{\pi}$.\newline

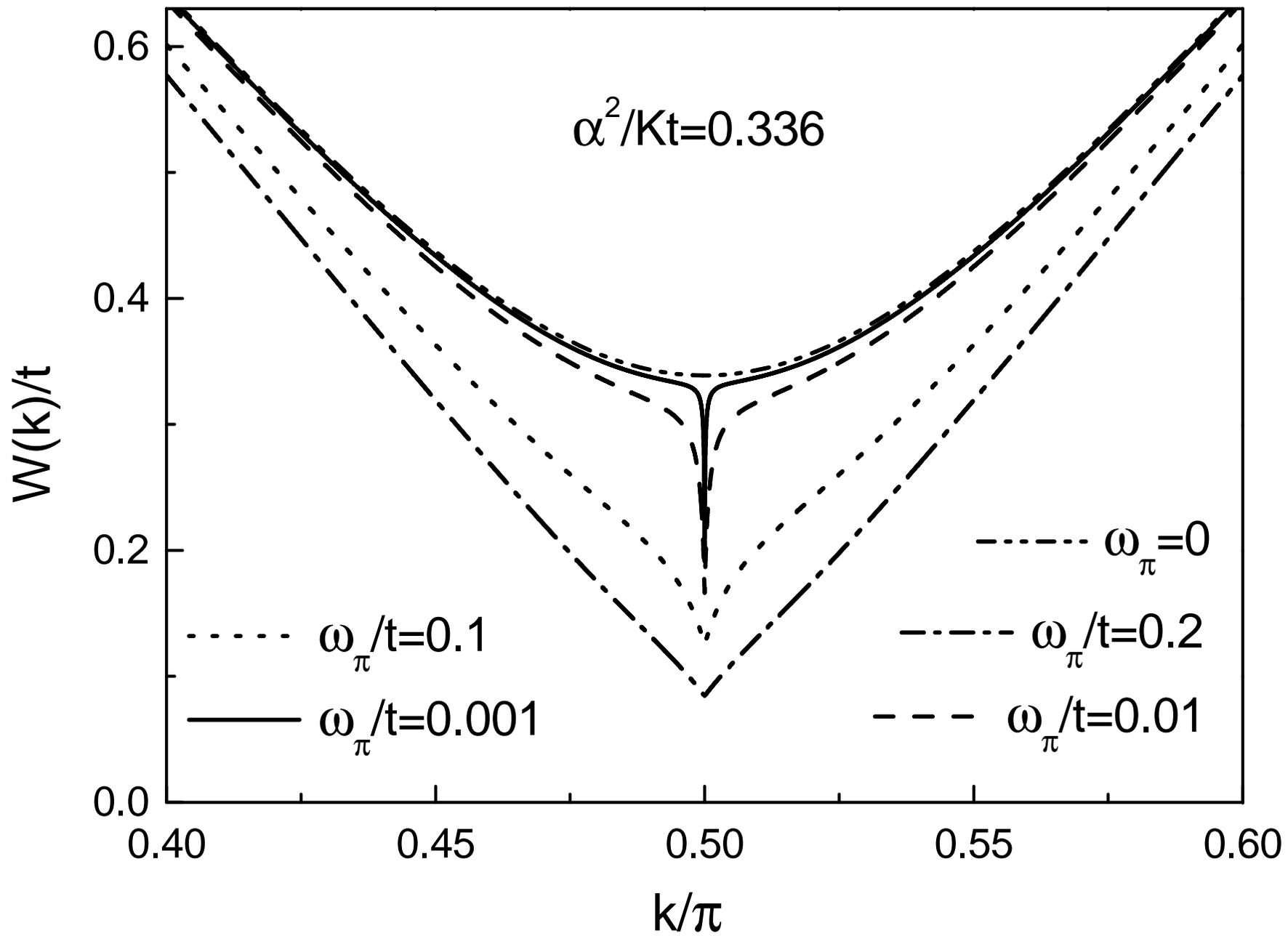
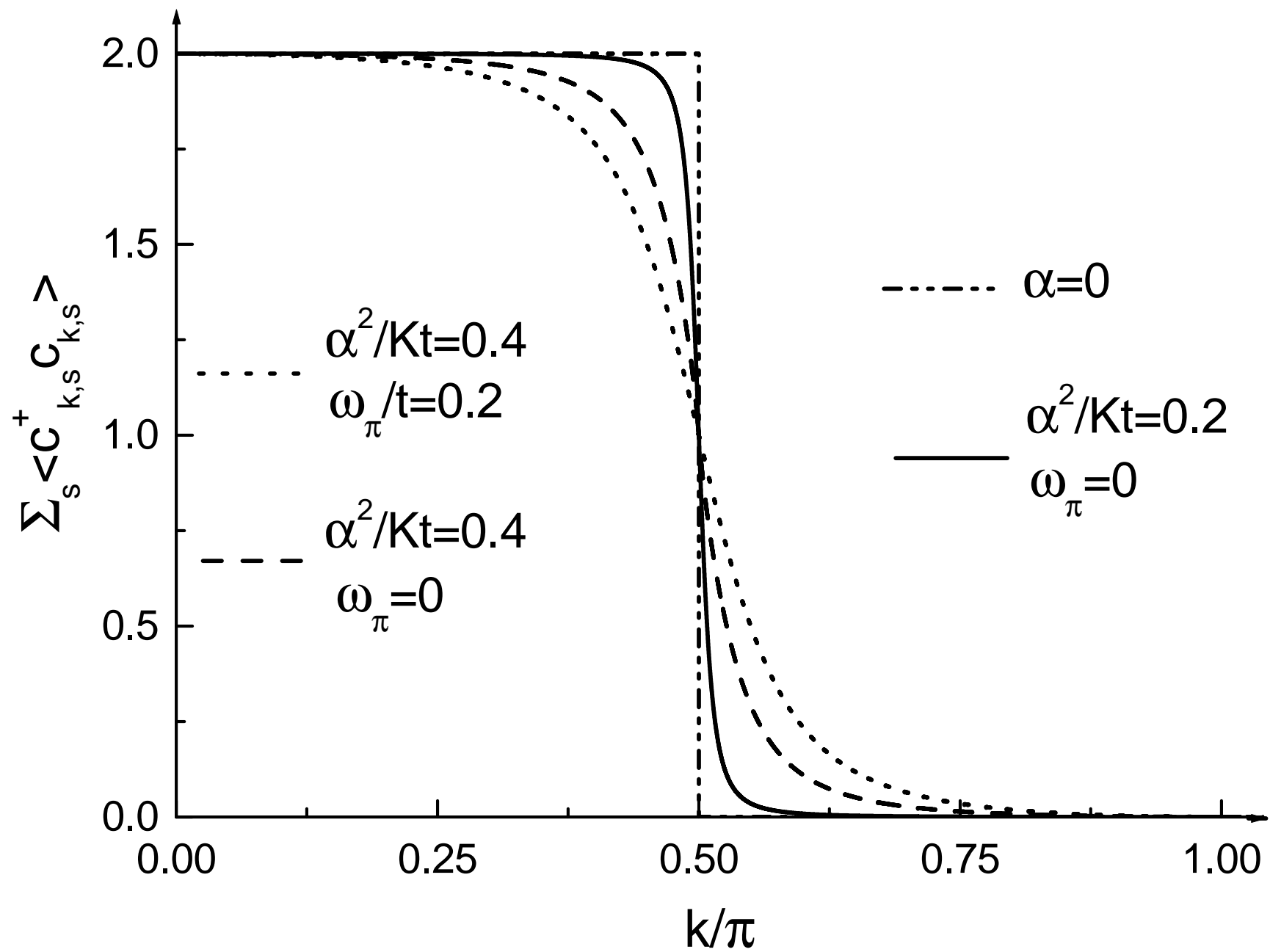
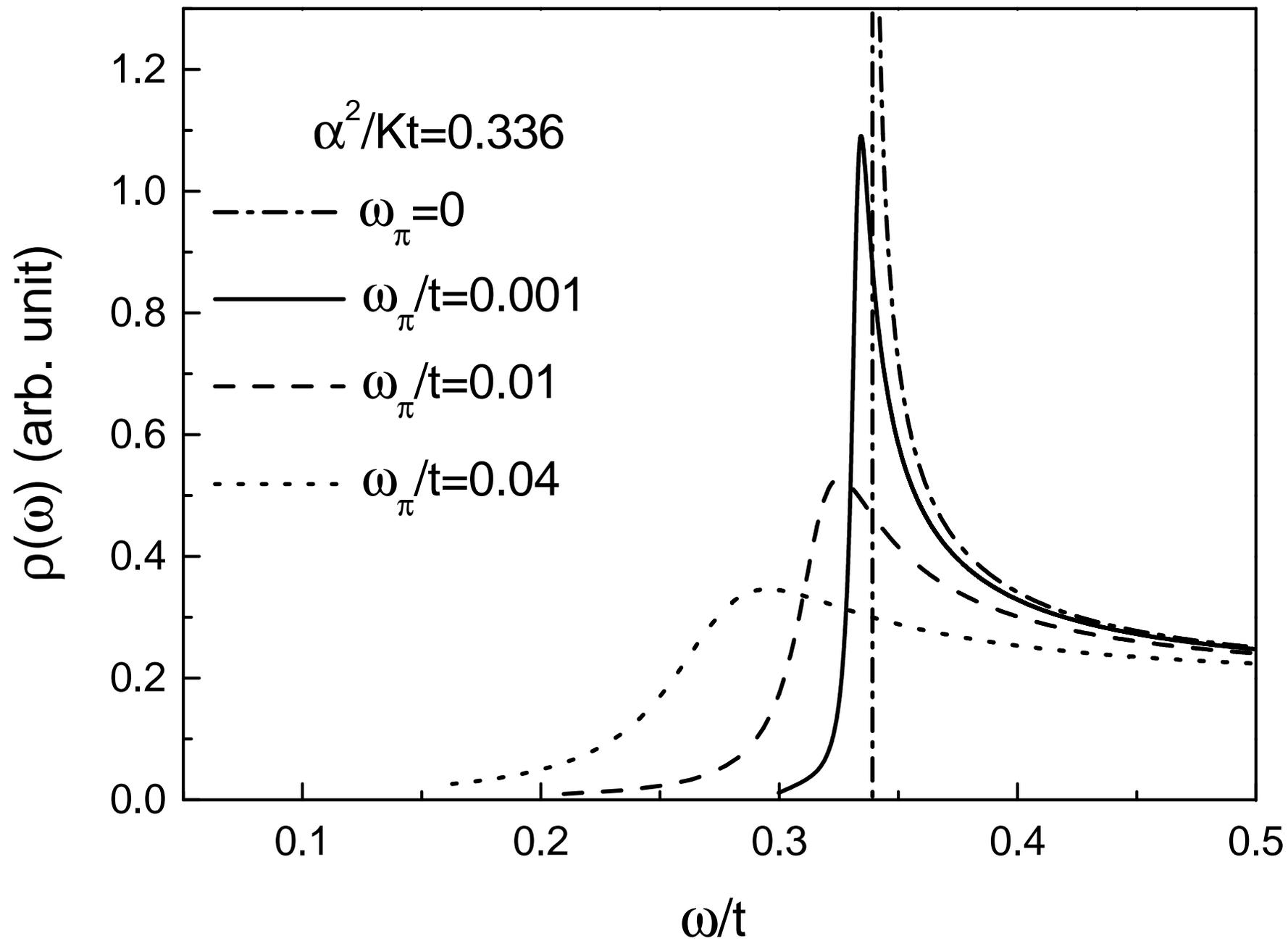
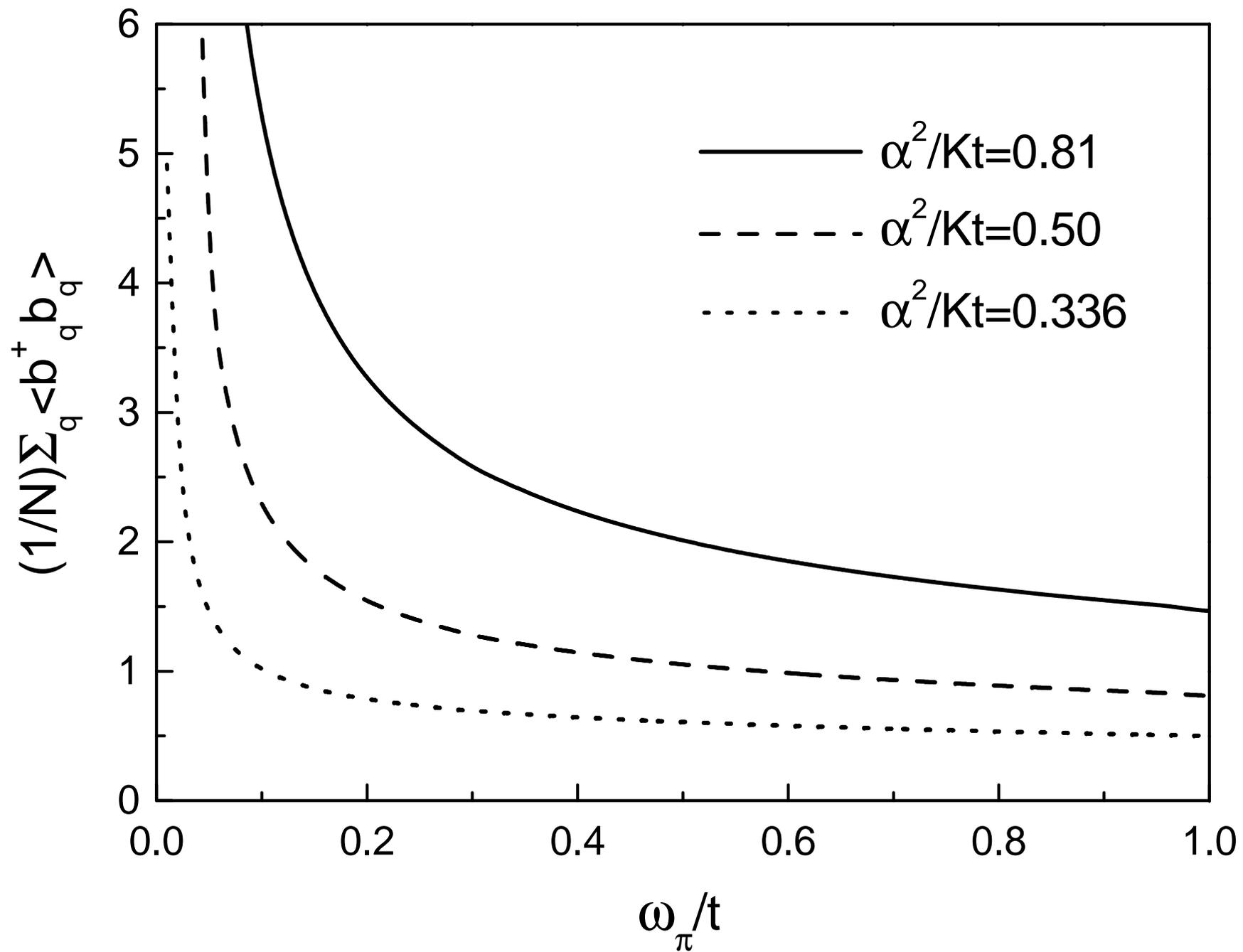
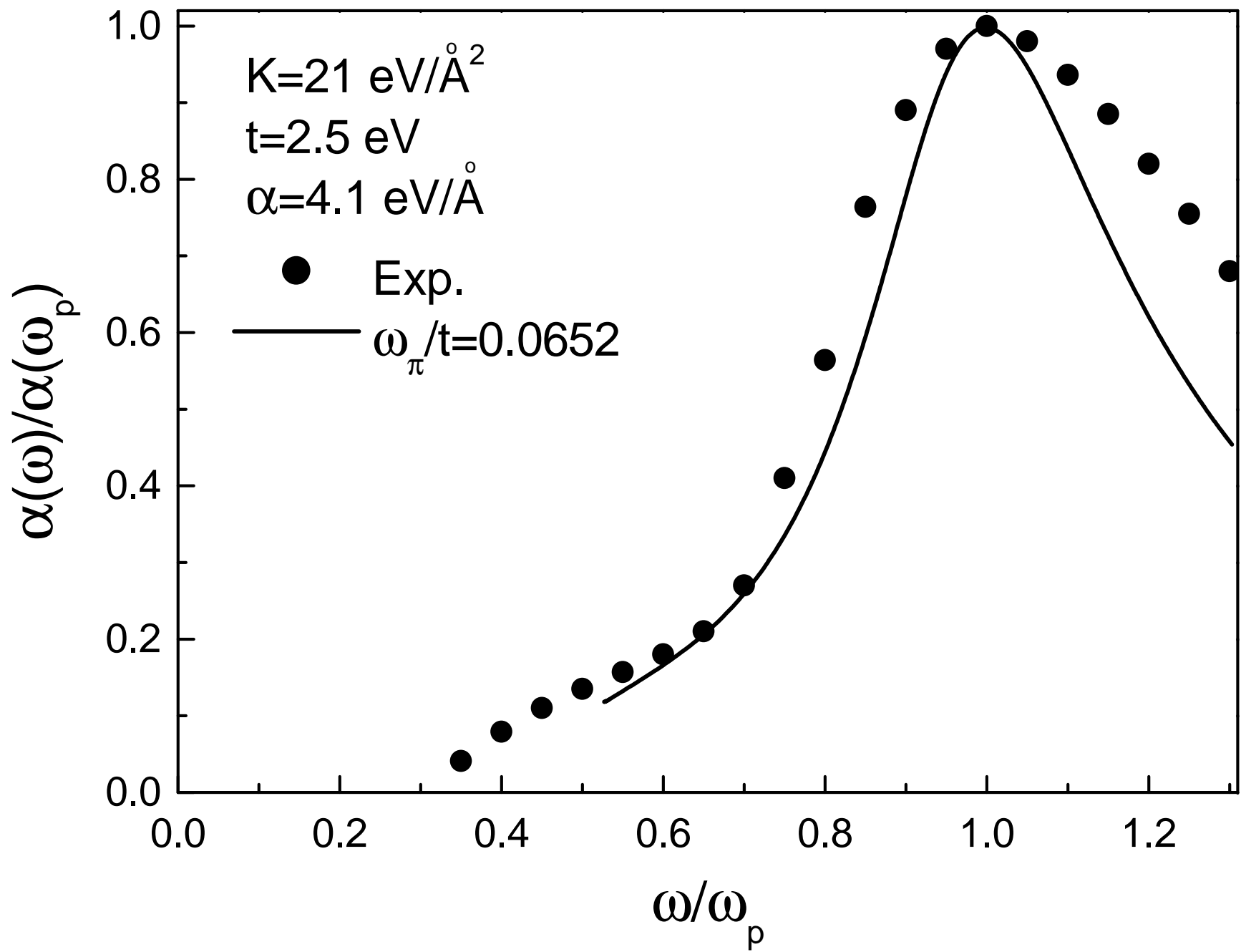
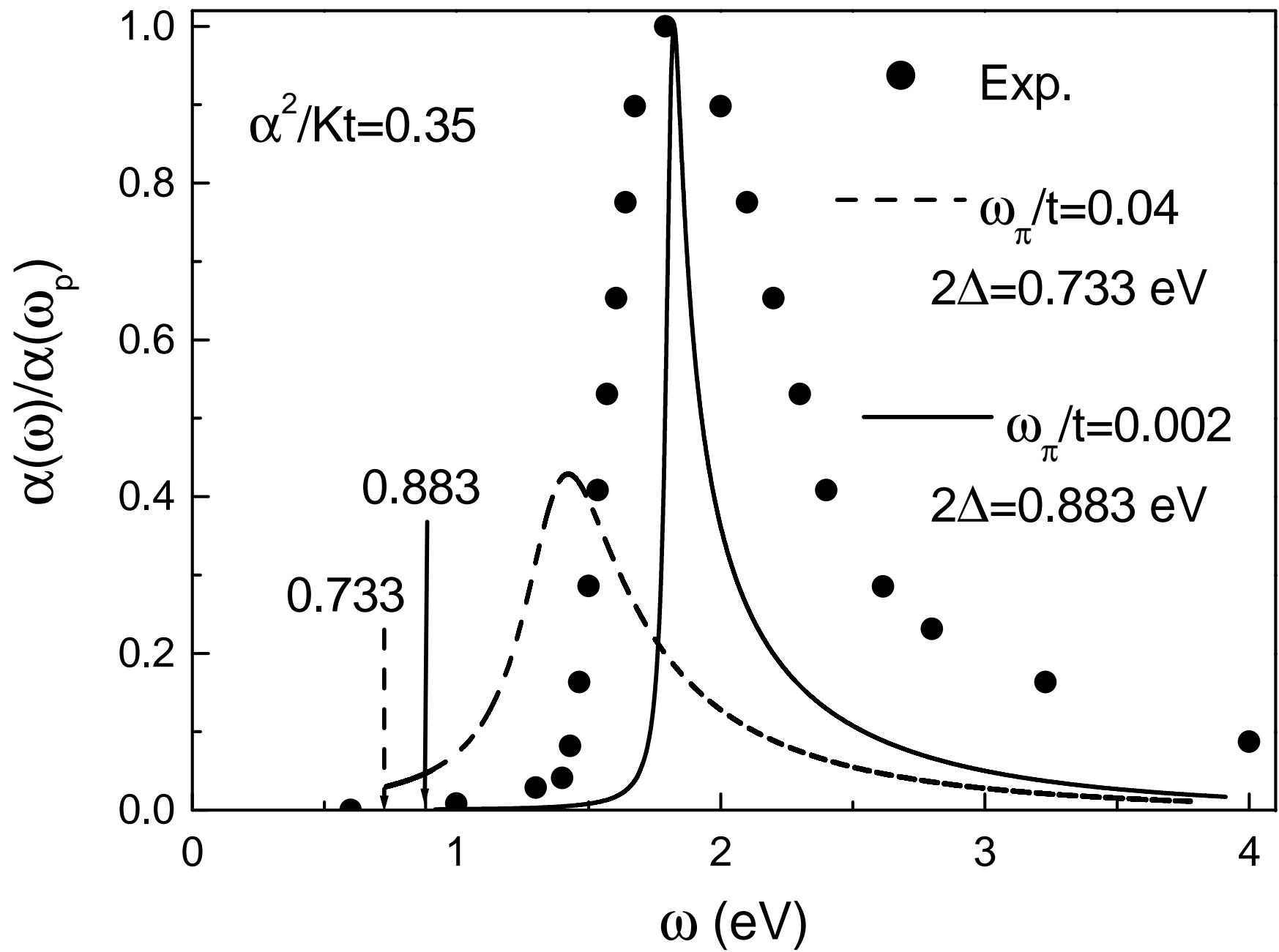
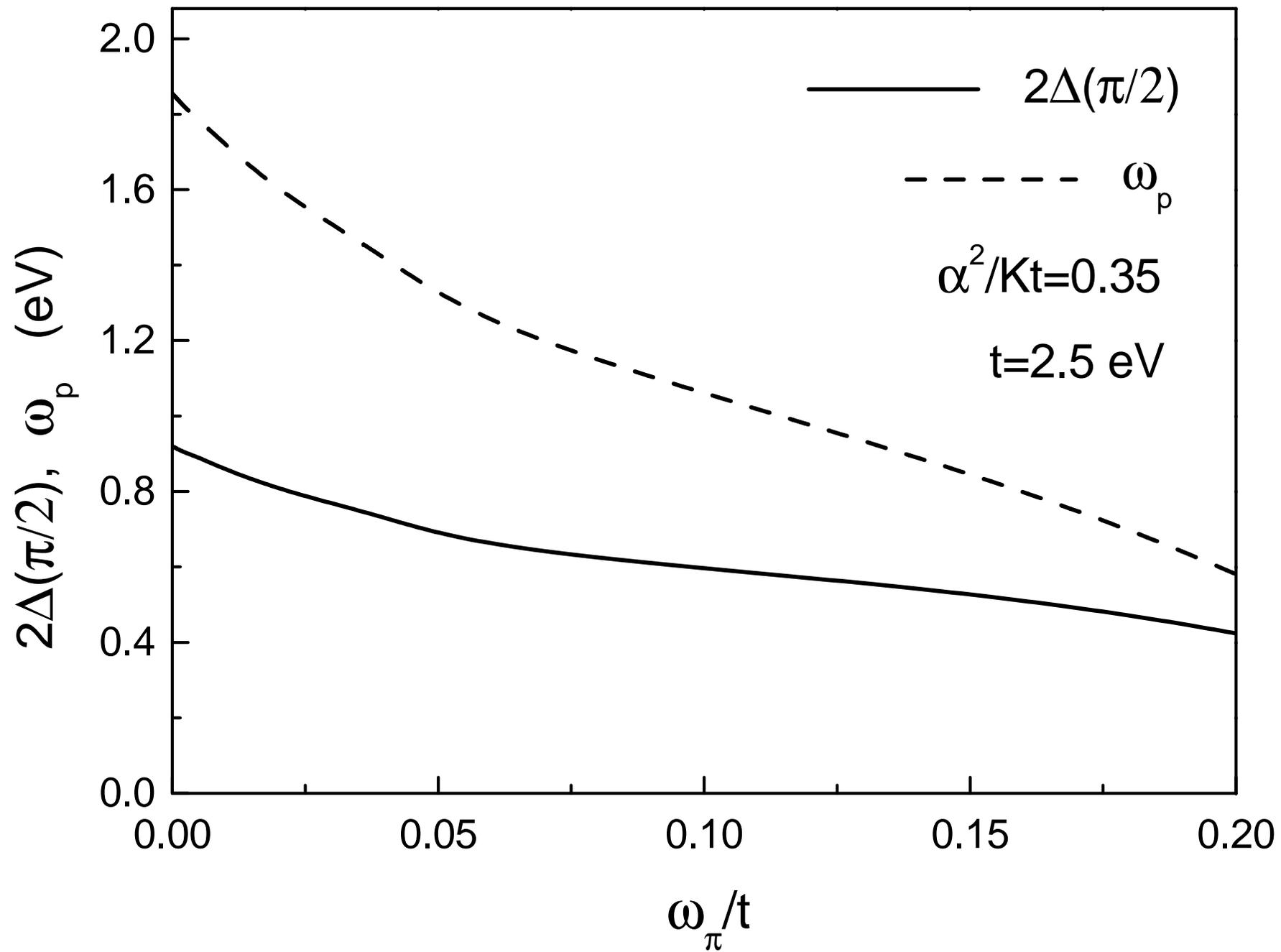
\end{document}